# Deep learning and classical computer vision techniques in medical image analysis: Case studies on brain MRI tissue segmentation, lung CT COPD registration, and skin lesion classification.

Daniel Tweneboah Anyimadu[a,1], Taofik Ahmed Suleiman[b,1,*], Mohammad Imran Hossain[c]

[a]Department of Computer Science, University of Exeter, Exeter, UK

[b]Wallace H. Coulter Department of Biomedical Engineering, Georgia Institute of Technology and Emory University, Atlanta, USA

[c]Department of Computer Science, Sorbonne Université, Paris, France

[1]Contributed equally to this work, [*]Corresponding author.

Email addresses: da536@exeter.ac.uk (DT Anyimadu), tsuleim@emory.edu (TA Suleiman), hossainimran.maia@gmail.com (MI Hossain)

**ABSTRACT**

Medical imaging spans diverse tasks and modalities which play a pivotal role in disease diagnosis, treatment planning, and monitoring. This study presents a novel exploration, being the first to systematically evaluate segmentation, registration, and classification tasks across multiple imaging modalities. Integrating both classical and deep learning (DL) approaches in addressing brain MRI tissue segmentation, lung CT image registration, and skin lesion classification from dermoscopic images, we demonstrate the complementary strengths of these methodologies in diverse applications. For brain tissue segmentation, 3D DL models outperformed 2D and patch-based models, specifically nnU-Net achieving highest Dice Coefficient (DSC) of 0.9397, with 3D U-Net models on ResNet34 backbone, offering competitive results (with DSC 0.8946). Multi-Atlas methods provided robust alternatives for cases where DL methods are not feasible, achieving average DSCs of 0.7267 (with majority voting) and 0.7255 (with weighted by mutual information). In lung CT registration, classical Elastix-based methods outperformed DL models, achieving a minimum Target Registration Error (TRE) of 6.68 ± 5.98 mm, highlighting the effectiveness of parameter tuning. HighResNet performed best among DL models with a TRE of 7.40 mm. For skin lesion classification, ensembles of DL models like InceptionResNetV2 and ResNet50 excelled, achieving up to 90.44%, and 93.62% accuracies for binary and multiclass classification respectively. Leveraging the One-vs-All method, DL attained accuracies of 94.64% (mel vs. others), 95.35% (bcc vs. others), and 96.93% (scc vs. others), while ML models specifically Multi-Layer Perceptron (MLP) on handcrafted features offered interpretable, efficient alternatives with 85.04% accuracy using SMOTE for class imbalance correction on the multi-class task and 83.27% on the binary-class task. This study represents the first comprehensive evaluation across segmentation, registration, and classification tasks, discussing methodologies for different imaging modalities, and emphasizing the importance of task-specific adaptations. Links to source code available on request.

## 1. Introduction

Medical imaging plays a pivotal role in understanding body structures and aiding diagnosis (Brammer, 2009; Shetewi et al., 2020; Siuly & Zhang, 2016). Recent advancements in artificial intelligence (AI), particularly in deep learning (DL) and machine learning (ML), have revolutionized medical imaging by enabling automation and precision in tasks such as segmentation (Andresen et al., 2022), registration (M. Chen et al., 2023), and classification (Litjens et al., 2017). While image registration aligns medical images for applications like atlas creation, intra-patient scan alignment, and volume quantification (Antoine Maintz, 1998), image segmentation, partitions images into distinct regions, extracting meaningful information vital for medical diagnostics and treatment planning (Andresen et al., 2022; Pham et al., 2000). Segmentation tasks, achieved through supervised or unsupervised methods (Aganj et al., 2018; J. Chen & Frey, 2020; Wang et al., 2022; Zemel, 2010), frequently rely on supervised approaches for accurate boundary and feature delineation, yielding reliable results in complex scenarios (Isensee et al., n.d.; Ronneberger et al., 2015). Similarly, image classification categorizes medical images into predefined classes, enabling tasks such as identifying pathological conditions, grading disease severity, and streamlining diagnostic workflows. Building on this foundation, we discuss the segmentation of brain MRI tissues, registration of lung CT COPD images, and classification of dermatological skin images.

Starting with the segmentation task, we utilized the IBSR18 dataset to explore and implement various approaches for brain MRI tissue segmentation. The objective was to generate accurate segmentation maps delineating white matter (WM), gray matter (GM), and cerebrospinal fluid (CSF) from MRI brain images, despite the dataset's inherent challenges of variable resolution and contrast. Addressing these complexities is vital for developing a robust segmentation model. This work aims to advance brain MRI tissue segmentation by investigating diverse data preprocessing techniques and segmentation models.

Continuing with the registration task, we focused on lung CT image registration in the context of chronic obstructive pulmonary disease (COPD), which is crucial for diagnosing and characterizing lung disorders (G. Wu et al., 2013; L. Zhang et al., 2022). Specifically, we explored the alignment of maximum inhale and exhale phase images, a key process in understanding respiratory dynamics. Utilizing the 4DCT DIR-Lab Challenge dataset (Castillo et al., 2013), we employed the Insight Segmentation and Registration Toolkit (ITK) (Martin et al., 2005) and the Elastix framework to perform precise lung CT image registration. Additionally, we compared these results with various deep learning-based registration algorithms from the Monai framework (Jorge Cardoso et al., 2022). This task aims to advance diagnostic accuracy, optimize treatment planning, and improve the monitoring of respiratory conditions.

The final task focuses on the classification of skin lesions, a challenging endeavor due to class imbalance and dataset bias. Class imbalance occurs when certain lesion types are overrepresented, leading to biased models and reduced accuracy for underrepresented classes and diverse demographic groups (Johnson & Khoshgoftaar, 2019). Addressing these challenges is essential to developing fair, accurate, and generalizable AI models in dermatology. To mitigate class imbalance, various methods have been employed. Techniques such as the Synthetic Minority Over-Sampling Technique (SMOTE) (Peranginangin et al., 2020) generate synthetic samples for minority classes, while data augmentation (Maharana et al., 2022; Shorten & Khoshgoftaar, 2019) enriches datasets through transformations like cropping, rescaling, contrast adjustments, zooming, and flipping. Other strategies, including cost-sensitive learning (Khan et al., 2018) and deep learning approaches using Convolutional Neural Networks (CNNs) (Bria et al., 2020), further enhance model performance by ensuring balanced representation across lesion types. Building on advancements in AI for skin lesion classification, this work develops robust and generalizable models leveraging both DL and ML techniques to address significant class imbalances effectively and improve classification outcomes across diverse lesion types and patient demographics.

In medical imaging, transfer learning-based deep learning (DL) has emerged as a powerful tool, enhancing model accuracy and efficiency by leveraging pre-trained networks such as InceptionResNet (Längkvist et al., 2014), ResNet (He et al., 2016), and MobileNet (Howard et al., 2017). Fine-tuning these models on specialized datasets enables the capture of complex patterns that traditional methods might overlook. In parallel, traditional machine learning (ML) approaches focus on feature engineering (Bro & Smilde, 2014; Heikkilä et al., 2009; Soh & Tsatsoulis, 1999) to analyze characteristics such as color, shape, and texture, further expanding the toolkit for imaging tasks. Recent studies highlight the effectiveness of AI in segmentation, registration, and classification. For instance, (Suleiman et al., 2024) achieved a balanced multiclass accuracy (BMA) of 91.07% by combining DenseNet121 with a random forest classifier to classify cancerous skin lesions. Similarly, (Mahbod et al., 2019) leverages Dense-Res-Inception Net (DRINet) in segmenting subtle features across medical images by combining dense connections, residual inception modules, and unpooling blocks. Additionally, (J. Zhang et al., 2019) introduce VoxelMorph as a fast, learning-based framework for deformable medical image registration, using a convolutional neural network to map image pairs to deformation fields. Such advancements demonstrate the potential of integrating DL and ML techniques to improve diagnostic workflows.

While segmentation, registration, and classification are often treated as separate tasks, they are interconnected in their contributions to advancing diagnostics, treatment planning, and overall healthcare delivery. This paper unites these key areas, presenting a comprehensive exploration of diverse classical and AI-driven methodologies. By bridging these independent projects, the study aims to provide a holistic understanding of their applications and impact on medical imaging, offering a unique resource for researchers and practitioners

## 2. Materials, and Methods

*2.1 Task 1: Brain MRI Tissue Segmentation*

The first task involves segmenting brain tissue into cerebrospinal fluid (CSF), white matter (WM), and gray matter (GM). The IBSR18 dataset was used for this project, consisting of 18 MRI volumes split into training (ten images), validation (five images), and testing (three images) sets as defined by the challenge organizers. An initial exploratory data analysis (EDA) was conducted to understand the characteristics of the IBSR18 dataset, revealing a significant class imbalance among the tissue categories. Notably, the CSF category, representing the minority class, accounts for less than 2% of the total ground truth segmentation volumes, as illustrated in Figure 1. This observation guided us in making informed decisions regarding preprocessing techniques to address the imbalance. For this task, we implemented both deep learning-based methods (Ronneberger et al., 2015) and classical methods based on Atlas (Cabezas et al., 2011). The approach was structured into three key steps: preprocessing, data extraction methods, and segmentation (deep learning and classical methods).

*2.1.1 Preprocessing of Brain MRI*

To address the heterogeneity in spatial resolutions and intensity distributions in the IBSR18 dataset, we applied Bias Field Correction (BFC) using the N4ITK algorithm (Tustison et al., 2010)(Kanakaraj et al., 2023) via the SimpleITK library (Yaniv et al., 2018). Otsu thresholding (Xu et al., 2011) was first used to create a binary mask, isolating brain tissues from the background. This enabled the N4ITK algorithm, recognized for its robustness in handling intensity inconsistencies, to focus corrections on relevant anatomical regions, avoid unnecessary adjustments to non-brain areas, and generate optimized corrected images for downstream segmentation tasks.

Prior to model training, pixel values were scaled to a common range of [0, 1] using min-max scaling (Patro & sahu, 2015). This normalization step is critical for managing the diverse spatial resolutions and pixel intensity ranges inherent in the dataset. It ensures consistency, particularly as the segmentation task focuses on CSF, GM, and WM, which exhibit distinct pixel intensity profiles. Figure 2 illustrates the preprocessing workflow, highlighting the effectiveness of each step in enhancing image uniformity for subsequent segmentation.

*2.1.2 Data Extraction Methods*

To facilitate the segmentation process and address the challenges posed by a limited training dataset and computational constraints for processing 3D data, we employed three distinct data extraction techniques. The first approach was patch-based training, which involved extracting non-overlapping 2D patches of a fixed size from the training images. These patches were then filtered to retain only those containing a substantial amount of non-zero pixels, ensuring the inclusion of relevant anatomical features. The second approach was slice-based training, which focused on extracting 2D slices from the training images. This method utilized a foreground mask to ensure that the extracted slices contained a significant number of non-zero pixels, thereby preserving essential spatial relationships and anatomical structures inherent in the 3D images. Additionally, to retain the originality of the 3D volumetric data, we incorporated the full 3D images into the training process. This approach complements the patch and slice-based methods by preserving the spatial continuity and volumetric context inherent in the dataset, supporting more robust model training. Figure 3 illustrates the results of these two extraction methods along with their corresponding labels, demonstrating the effectiveness of these techniques in preparing data for segmentation.

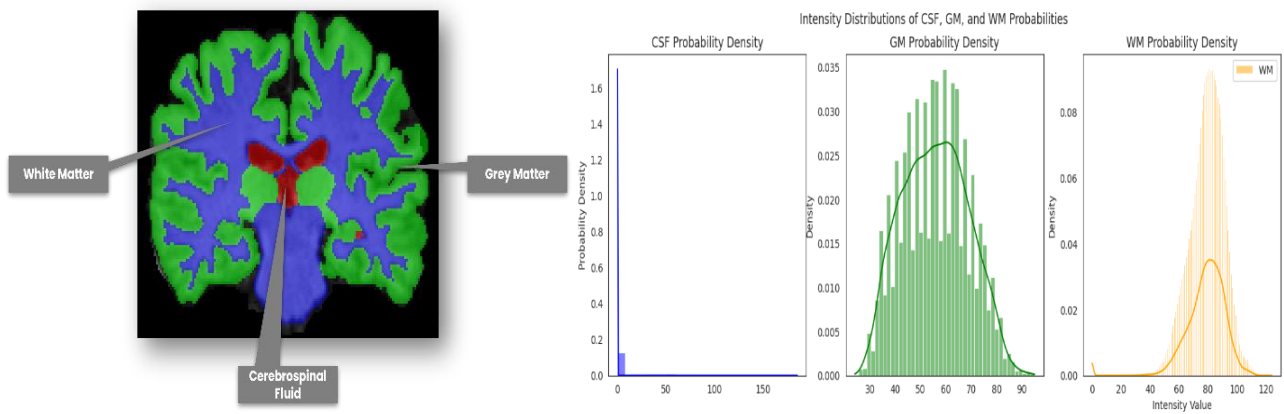

**Fig. 1.** Visual representation, and intensity distributions of the classes

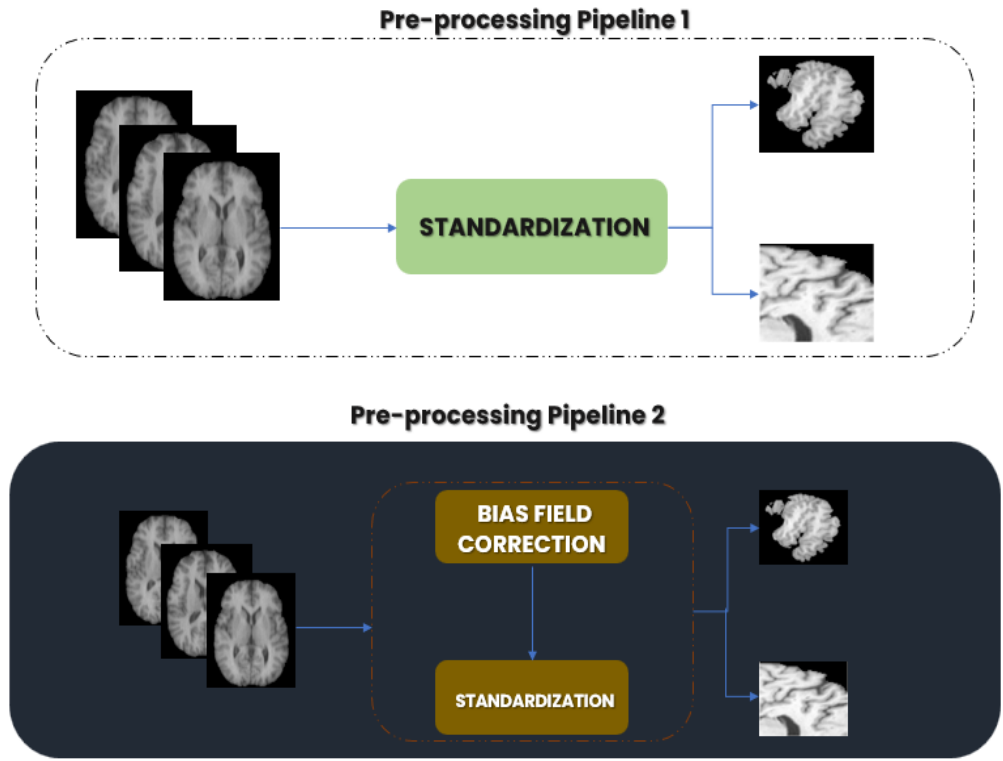

**Fig. 2.** Pipelines for the preprocessing techniques.

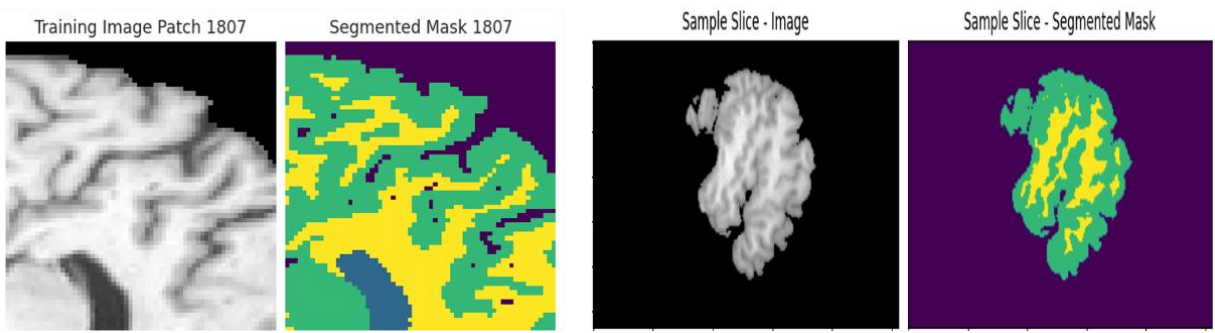

**Fig. 3.** Visualization of the 2D patches and slices

*2.1.3. Segmentation Using Deep Learning Based Methods*

For the deep learning methods, we implemented state-of-the-art segmentation models, including customized U-Net architectures and pre-trained backbones integrated into U-Net. Specifically, we explored the use of ResNet34 (Z. Wu et al., 2019), ResNet50 and InceptionResNetV2 (Längkvist et al., 2014) as backbones, along with a custom-built U-Net without a backbone. The custom U-Net architecture consists of a contraction path (encoder) and an expansive path (decoder), designed to capture both local and global features, making it well-suited for segmenting complex brain structures. The contraction path includes three convolutional blocks, each with two consecutive 3x3 filters using ReLU activation and "same" padding, followed by dropout layers to mitigate overfitting. Down-sampling is achieved using 2x2 MaxPooling2D layers. The expansive path mirrors the contraction path but uses Conv2DTranspose layers for up-sampling. Skip connections ensure seamless feature flow between the contraction and expansive paths. The output layer uses 1x1 filters with softmax activation to generate the final segmentation map, assigning class probabilities. Additionally, we used the monai U-Net architecture (Jorge Cardoso et al., 2022), and equally employed the nnU-Net (no-new U-Net) framework (Isensee et al., n.d.), an automated deep learning solution tailored for medical image segmentation tasks, enhancing our results further. Figure 4 illustrates the schematic representation of segmenting CSF, GM, and WM using a deep learning approach.

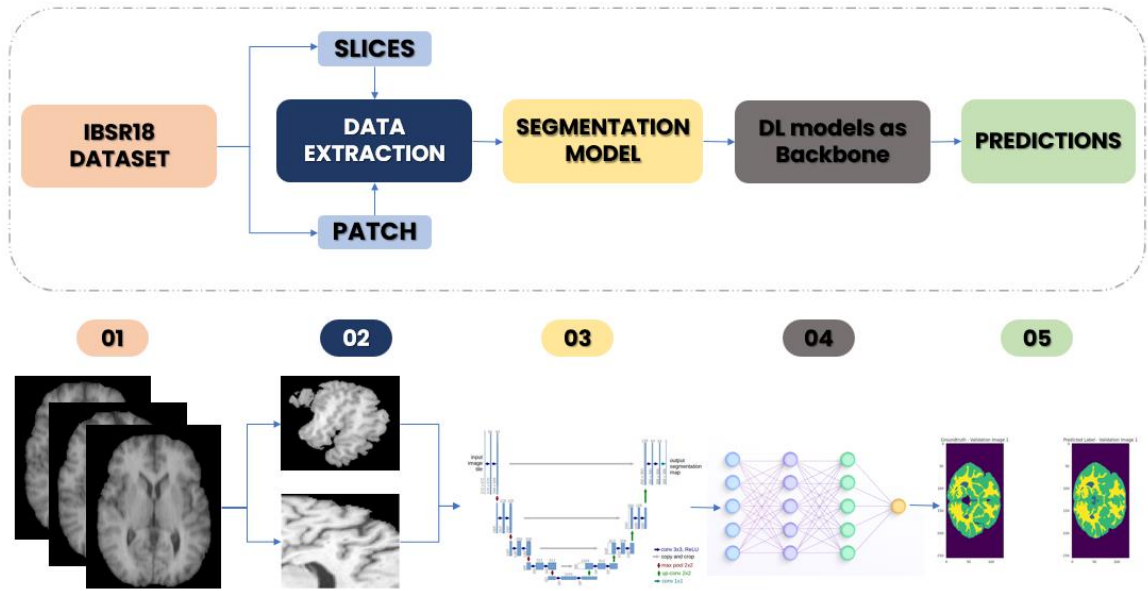

**Fig. 4.** Schematic diagram representing the segmentation of CSF, GM and WM via a DL approach

*2.1.4. Segmentation Using Atlas-Based Methods*

      For Atlas-based segmentation, we leveraged the power of registration to build a reference model capturing the spatial distribution of brain tissues. The probabilistic Atlas and Multi-Atlas methods were implemented. The probabilistic Atlas construction involves aggregating information from registered training images. For each tissue class (CSF, GM, WM), voxel intensity histograms were computed to create probability density functions (PDF). A topological atlas, representing the most probable tissue class at each voxel, was derived from these probabilistic atlases. Tissue segmentation involved creating probability atlases and a mean volume accumulator from the reference image. Intensities were normalized, and probabilistic atlases were constructed for each training image, yielding a topological atlas. Intensity histograms were computed for each tissue class using these atlases, with PDF and posterior probabilities calculated and visualized as shown in Figure 5 and 6 respectively.

      Label propagation was also employed, registering images using affine transformations and propagating labels to achieve precise segmentation. In the multi-Atlas approach, multiple atlases were generated by using each training image to create an atlas for every image in the validation set. Two label fusion methods were applied: majority voting and weighting by mutual information, ensuring a robust segmentation output. These methods together provide a comprehensive approach for accurately segmenting brain tissues into CSF, GM, and WM.

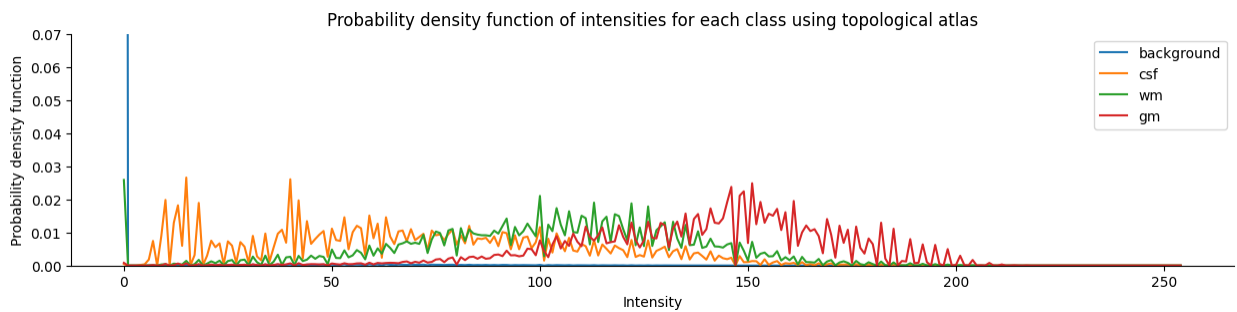

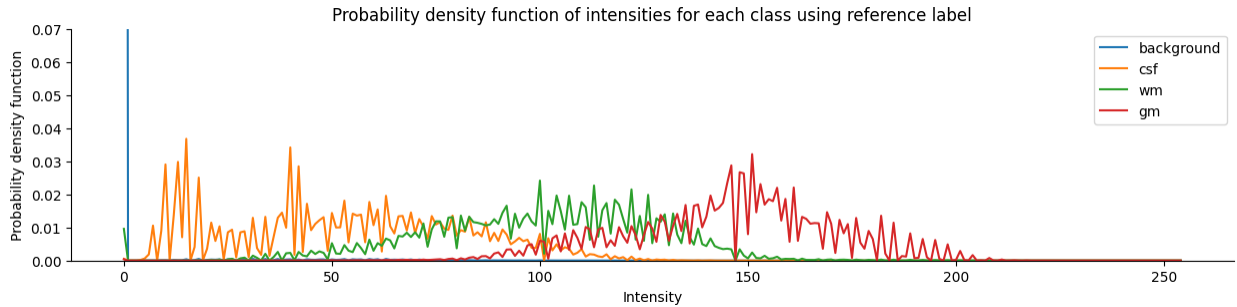

**Fig. 5.** Probability density functions (Topological atlas and reference label respectively)

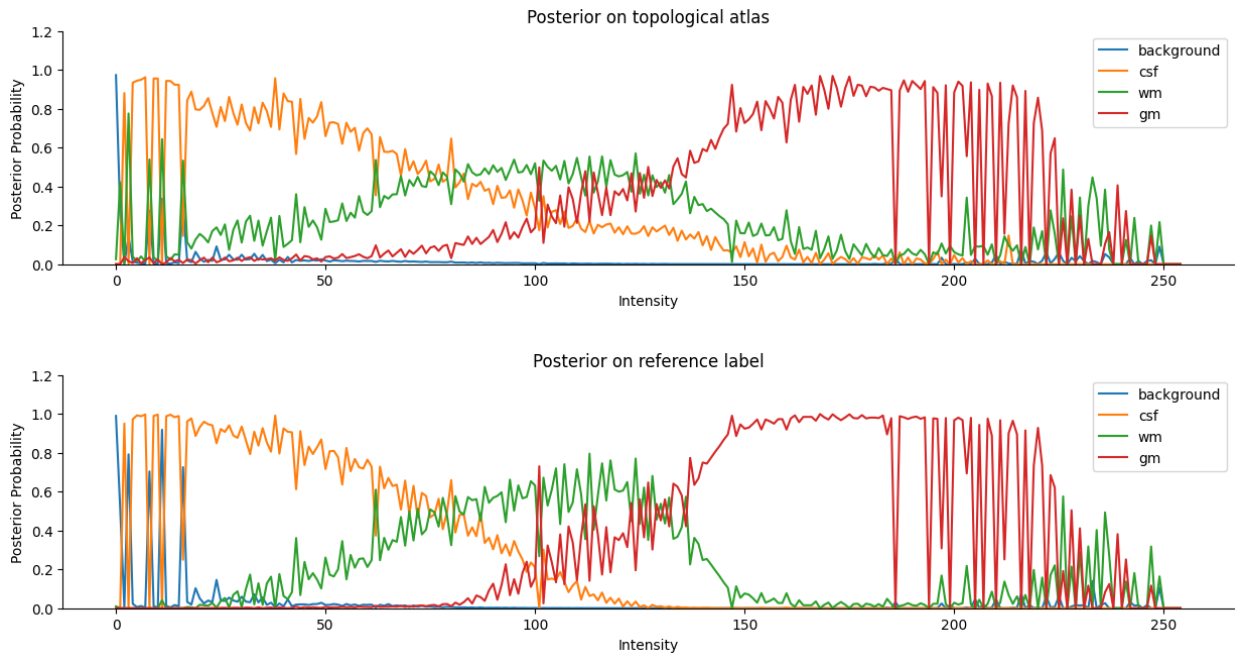

**Fig. 6.** Posterior (Topological atlas and reference label respectively)

*2.2 Task 2: COPD Lung CT Image Registration*

The second task is exploring several registration methods for lung CT registration which has been achieved in three key pipelines. These are data handling and processing, CT lung preprocessing, and finally, registration using either any of the pre-trained deep learning models (SegResNet, and HighResNet) from MONAI or Elastix/Transformix. Figure 7 shows a schematic diagram representing the registration of lung CT images (exhale onto inhale).

*2.2.1. Data Handling and Processing*

The dataset utilized is the 4DCT DIR-Lab Challenge dataset, comprising 10 pairs of lung CT scans. For this study, we focused on cases 1 to 4 for training and validation, with each pair containing maximum inhalation and exhalation scans. The dataset includes raw images and landmark files with 300 coordinates and mean displacement values. Since the data was in a raw format, it was essential to convert it into ITK-readable formats, such as NIfTI, to enable seamless integration with ITK and Elastix frameworks. To achieve this, we employed SimpleITK, which effectively handles raw binary scalar images. The process involved decoding the raw binary images and transforming them into the SimpleITK format, ensuring compatibility with the registration tools.

An exploratory data analysis was conducted to understand the characteristics of the 4DCT DIR-Lab Challenge dataset. By visualizing the images alongside their anatomical landmarks, we identified precise landmark locations critical for accurately aligning inhale and exhale images, thereby facilitating detailed feature comparisons. During this exploration, we encountered a notable challenge: the lungs were embedded in a circular gantry region (CT table), complicating the distinction between the intensity distribution and spatial locations of the inhale and exhale images. This obstruction would hinder registration without removing the CT table. Consequently, this challenge informed our preprocessing strategy to leverage the dataset effectively and ensure the table's removal. Figure 8 illustrates the CT lung images during exhalation and inhalation at various positions with their corresponding landmarks.

*2.2.2. CT Lung Preprocessing*

To effectively prepare the CT images for subsequent registration with both ITK Elastix and deep-learning frameworks, we developed a robust preprocessing class to remove the table. This comprehensive class incorporates multiple key functions designed to enhance image quality and ensure compatibility with registration workflows. The preprocessing begins with an evaluation of the field of view (FOV), which examines intensity levels in specific corners of the image. If the intensity in at least three corners falls below a specified threshold, it indicates the presence of FOV. Based on this assessment, the subsequent segmentation process is adjusted, with a K value of 3 if FOV is present and 2 otherwise. This step ensures an accurate initialization of the segmentation pipeline.

Following FOV evaluation, the images undergo segmentation using the K-Means clustering algorithm (Ikotun et al., 2023). The inverted image is vectorized and clustered into distinct groups, and the resulting centroids are utilized to reconstruct the segmented image. This segmentation step is crucial for isolating anatomical structures from the table and facilitating the removal of artifacts. To address the presence of holes or gaps in the segmented image, a chest hole-filling function is applied. This step employs contour detection to identify the contour with the maximum area, which is then used to create a mask for filling the corresponding holes. Morphological operations (Comer & III, 1999) were used to refine the mask, ensuring a complete and accurate representation of the lung structure.

Subsequently, the segmented image and contours from the chest hole-filling process is used to eliminate gantry artifacts, including remnants of the CT table. This is done by multiplying the input image with the refined contours, this step effectively removes unwanted elements, significantly improving the quality of the images. Finally, to enhance image contrast and highlight subtle anatomical features, the Contrast Limited Adaptive Histogram Equalization (CLAHE) algorithm is applied. This step prevents over-saturation in homogeneous regions and contributes to better visualization of important details, crucial for precise registration and further analysis. Figure 9 illustrates the general workflow of these preprocessing steps, demonstrating the seamless transition of data between each stage. Figure 10 presents intermediate results and the final output, showcasing the effectiveness of this preprocessing pipeline in enhancing CT images for the subsequent registration process

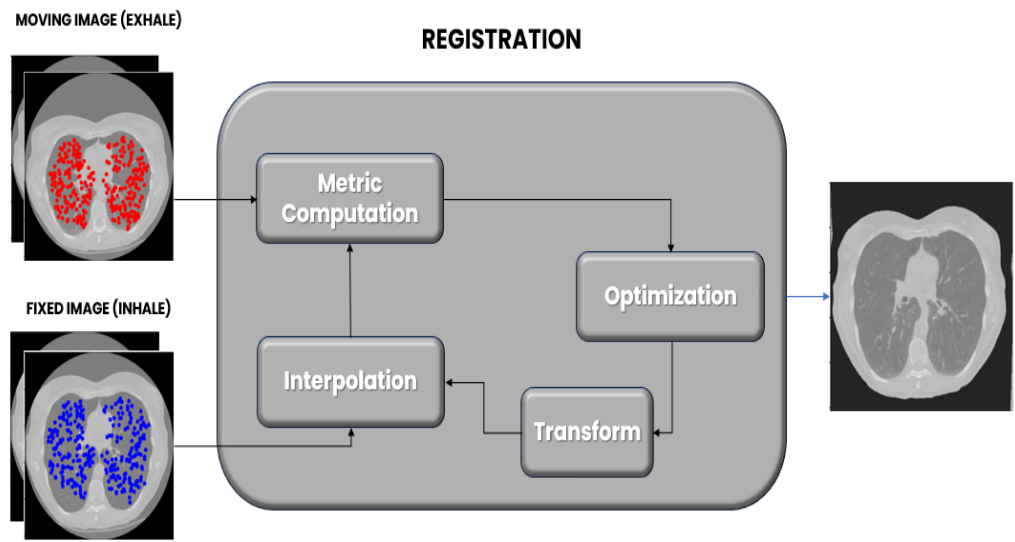

**Fig. 7.** A schematic diagram representing the registration of lung CT images (exhale onto inhale).

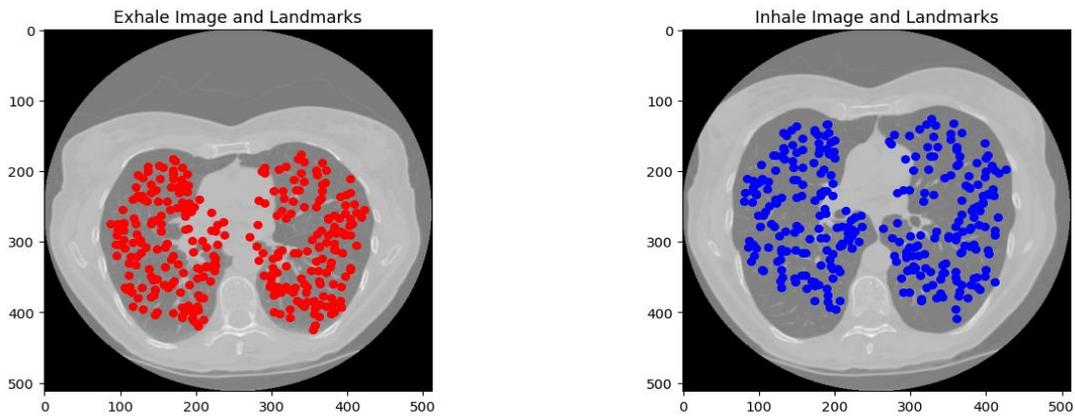

**Fig. 8.** Original lung CT images with their landmark points, indicated by red and blue dots

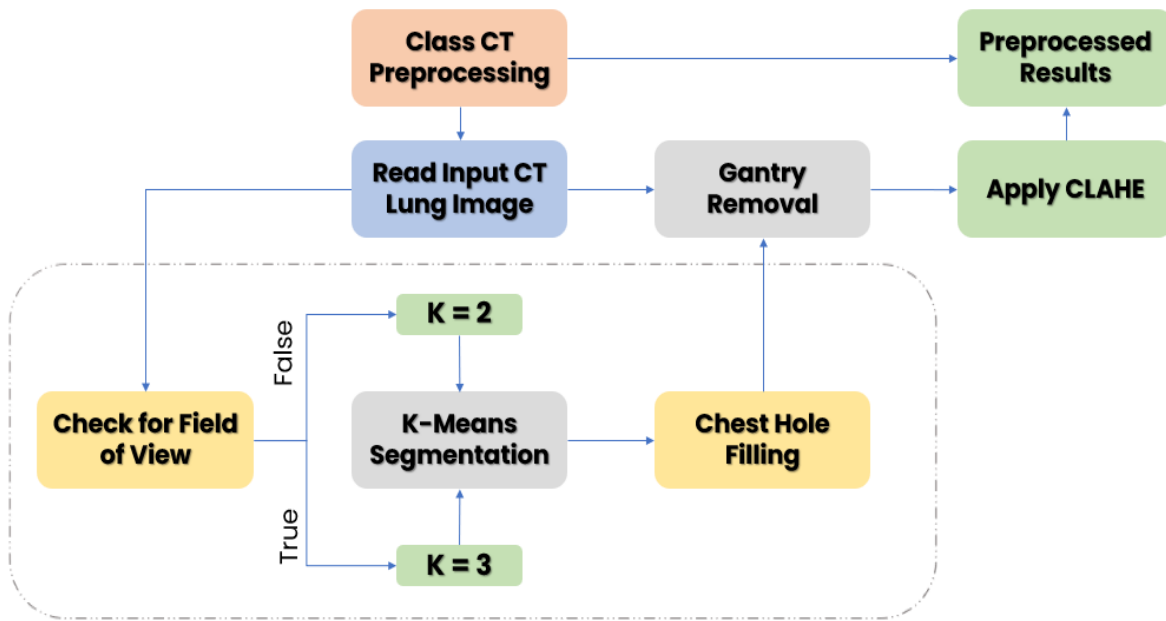

**Fig. 9.** General workflow of the functions within the class, showcasing the seamless transition of data between each processing step.

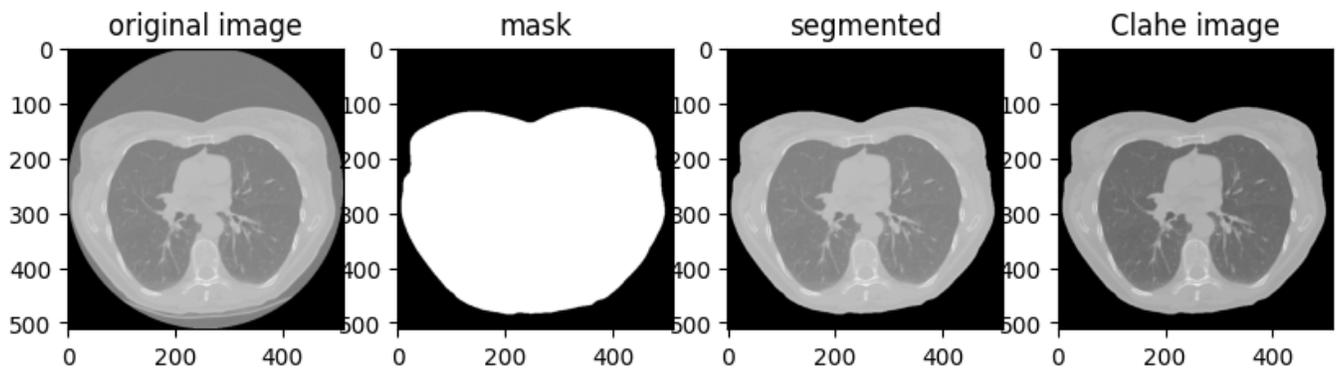

**Fig. 10.** Visualizations of the original image, segmentation mask, segmented image after gantry removal, and the final CLAHE-enhanced image.

*2.2.3. Registration Using Deep-Learning Based Methods*

The second registration method employed advanced deep learning (DL) models, leveraging pre-trained architectures within the MONAI library to register lung CT images with associated landmarks. A significant challenge was preparing the data in a format suitable for training deep learning registration models with landmarks. To address this, we utilized preprocessed images and developed a custom data loading function to efficiently organize images into training and validation sets within a dictionary structure. This dictionary included keys for fixed and moving images, as well as fixed and moving landmarks.

Given that MONAI's transforms are primarily designed for images with an emphasis on spatial transformations, adapting them for point clouds presented unique challenges. To overcome this, we implemented two custom classes for loading and transforming keypoints. These classes ensured that linear transformations were applied consistently with the underlying image transformations, such as affine transformations used during augmentation. Figure 11 showcases an example visualization of images and their corresponding point clouds.

The hyperparameters were configured to optimize the target registration error (TRE) loss, focusing on point-based accuracy. The forward function predicted the Displacement Field (DDF) and warped the moving image accordingly, while the collate function ensured the alignment of key points for batch processing. The loss function incorporated multiple components, including TRE, Mean Squared Error (MSE), and Bending Energy, to optimize model performance during training. The models trained are SegResNet, and HighResNet. The training process involved model initialization with pre-trained architectures, optimization using the Adam optimizer with a cosine annealing learning rate scheduler, and a training-validation loop where TRE was the primary loss metric. The best-performing models were saved for future predictions. Each model has variations in architecture configurations, for example SegResNet included a warp layer for image warping based on the DDF and incorporated TRE, MSE, and Bending Energy in the loss function and HighResNet incorporated padding for channel matching during network operations.

*2.2.4. Registration Using Elastix and Transformix Methods*

The classical registration process begins with Elastix, using input parameters such as the fixed image (inhale), moving image (exhale), transformation parameters, and an output directory. This process transforms the exhale images to align with their corresponding inhale images, producing transformed exhale images and their associated transformation parameter results. Following Elastix, Transformix is applied, which uses the moving image (exhale), the landmarks of the inhale image, and the transformation parameter results from Elastix to generate landmarks for the exhale images.

To refine the registration process and achieve more robust results, we experimented with parameter maps curated specifically for Chest/Lung 3D/4D CT image registration, available in the Elastix-ModelZoo documentation. These parameter maps provide flexibility through options such as multi-resolution registration, image pyramid schedules, and customizable components, including interpolators, optimizers, metrics, and iteration settings. The inbuilt Affine and B-spline transformations from ITK served as initial benchmarks for the registration process. After a comprehensive review of parameter maps, we selected three promising configurations: Par0007 (intrapatient B-spline transformation using mutual information), Par0011 (intrapatient B-spline transformation with normalized correlation), and Par0049 (intra-subject B-spline transformation using Mattes mutual information). To optimize outcomes, we combined multiple parameter maps and customized specific features, such as image pyramid schedules, to enhance performance. The final strategy involved combining the best-performing parameter files to achieve optimal registration results.

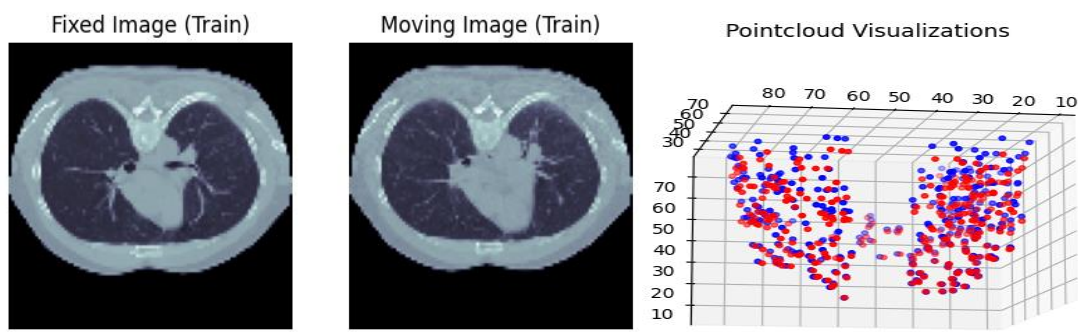

**Fig. 11.** Data visualization from the deep learning data and landmark loader

*2.3 Task 3: Skin Lesion Classification*

The final task focuses on the classification of skin lesions using both deep learning (DL) and machine learning (ML) approaches. Images from the HAM10000 (ViDIR Group, Medical University of Vienna) (Tschandl et al., 2018), BCN_20000 (Hospital Clínic de Barcelona) (Hernández-Pérez et al., 2024), and MSK (ISBI 2017) (F Codella et al., 2017) datasets were utilized, encompassing six skin lesion types. A subset of these datasets was selected to include a variety of cases for this study. Two classification tasks were performed: binary classification and multi-class classification. For the binary classification task, the dataset included 7725 nevus images and 7470 images of other lesion types, such as pigmented benign keratoses, dermatofibroma, melanoma, and basal cell carcinoma. For the multi-class classification task, the dataset contained 2713 melanoma, 1993 basal cell carcinoma, and 376 squamous cell carcinoma images. Each image was labeled with its corresponding lesion type to enable supervised learning. Illustrations of different lesion types are shown in Figure 12.

Before beginning the classification tasks, EDA was conducted to visualize and understand the dataset's characteristics, such as class distributions and image features. The EDA revealed significant class imbalances, particularly in the multi-class classification task, where squamous cell carcinoma (SCC) images were underrepresented. These insights guided the selection of strategies such as SMOTE, data augmentation, and class weighting to address the class imbalance and improve model generalizability. The methodology for this task was structured into three key parts: addressing class imbalance, feature extraction and engineering for the ML approach, and classification using both DL and ML methods.

*2.3.1 Data Augmentation, and Class Imbalance Techniques*

To address the class imbalance inherent in the skin lesion datasets, various data augmentation (Maharana et al., 2022; Shorten & Khoshgoftaar, 2019) techniques were employed to artificially increase the diversity and quantity of samples from minority classes. For the ML approach, augmentation methods such as central cropping, rescaling, and contrast enhancement were applied, while DL models utilized zoom range and horizontal/vertical flipping to generate augmented images, particularly for minority classes like basal cell carcinoma (BCC) and squamous cell carcinoma (SCC). These techniques were designed to balance the dataset and enhance learning in both binary and multi-class classification tasks, ensuring that the augmented data remained relevant without introducing noise or irrelevant features.

The SMOTE was also used in the ML approach to increase the representation of minority classes (BCC and SCC) in the training data. SMOTE operates by generating synthetic samples through interpolation between existing minority class samples, effectively expanding the dataset while mitigating the risk of overfitting to the majority classes. The synthetic data closely resembled real samples, ensuring that the models were trained on data distributions representative of clinical scenarios, thereby improving generalizability.

For DL models, class weights were calculated and applied during training to address the class imbalance. This involved assigning higher weights to minority classes like BCC and SCC, ensuring their influence on the loss function (Kumar, 2022) was proportionate to their clinical importance despite lower representation. By incorporating class weights, the DL models were encouraged to focus adequately on these underrepresented classes, preventing overfitting to majority classes and enhancing overall prediction balance. This strategy was critical for achieving more equitable performance, particularly in clinical applications where misclassification of rare lesion types could have significant consequences.

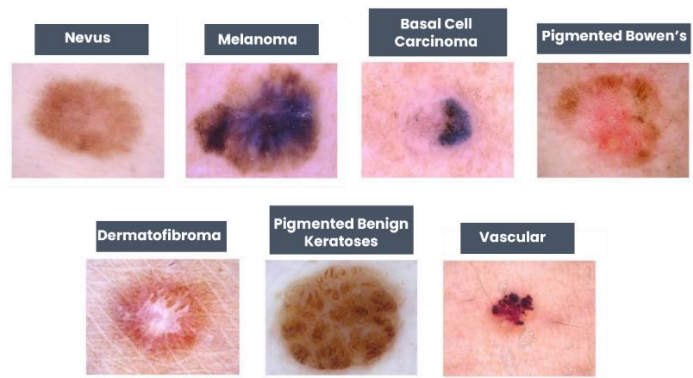

**Fig. 12.** Illustrations of different lesion types

*2.3.2 Feature Extraction and Engineering*

To classify skin lesions effectively using machine learning, we focused on extracting features based on asymmetry, border, color, diameter (ABCD rule), and texture patterns (Kestek et al., 2023)(Vocaturo et al., 2019). For color features, we calculated key statistics for each channel in both RGB and HSV color spaces, including mean, standard deviation, minimum, maximum, skewness, kurtosis, and entropy. Shape features were extracted using methods like Canny Edge Detection (Xuan & Hong, 2017) and Hu Moments (R. Zhang & Wang, 2011) to capture the morphological characteristics of skin lesions. The Canny Edge Detection method identified edges by employing a Gaussian filter to suppress noise, Sobel filters to calculate gradients and directions, and non-maximum suppression to thin the edges. Meanwhile, Hu Moments, invariant to translation, rotation, and scale, were computed from central moments to provide a compact and robust representation of shape, capturing properties like perimeter, area, circularity, and compactness.

For texture features, we employed the GLCM and LBP. GLCM was used to analyze the spatial relationship between pixel intensities, extracting features such as contrast, homogeneity, energy, and correlation, using distances (2, 5, 7, 10, 15) and angles (0, 45, 90, 135). LBP, on the other hand, captured local texture patterns by comparing each pixel with its neighbors to form binary patterns representing texture. Radii ranging from 1 to 9 were used, with points calculated as 8 x radius, providing detailed texture descriptions. Together, GLCM and LBP offered a comprehensive analysis of texture, critical for distinguishing between various skin conditions.

To ensure these features were optimal for classification, feature scaling and selection techniques were applied. Standardization was used to normalize the features, transforming them to have a mean of 0 and a standard deviation of 1, ensuring consistency across the dataset. For feature selection, we employed PCA, retaining the top 70 components. These techniques helped focus on the most relevant features, enhancing the model's performance by reducing dimensionality and emphasizing key attributes for classification.

*2.3.3 Classification Using Deep-Learning Based Methods*

We utilized various DL models, each selected for their proven effectiveness in skin lesion classification tasks. The models included DenseNet121/210 (DNET) (Zhu & Newsam, 2017), ResNet (He et al., 2016), MobileNet (Howard et al., 2017), and InceptionResNetV2B0 (IRNET) (Längkvist et al., 2014). DenseNet121/210 leverages dense connectivity and efficient parameter sharing to identify intricate patterns and features essential for distinguishing between skin lesion types. ResNet50 uses skip connections within a residual learning framework, addressing the vanishing gradient problem and capturing detailed features in heterogeneous skin lesion images. MobileNet employs depth-wise separable convolutions, which reduce parameters and computational cost while maintaining high classification accuracy, making it particularly suitable for diverse lesion types and large datasets. InceptionResNetV2B0 combines the Inception architecture with residual connections, creating a robust model capable of capturing a wide range of features. Additionally, for the three-class classification problem, we implemented a one-vs-all strategy using IRNET.

Prior to training the DL models, we employed a stratified class-wise splitting strategy using the Stratified Shuffle Split method. This approach allocated 80% of the dataset for training and 20% for cross-validation, ensuring consistent class distributions and reliable validation metrics. Pre-trained architectures were adapted for training by freezing the initial layers, which were pre-trained on ImageNet, and adding custom layers on top. These custom layers included configurations such as Global Average Pooling, Flatten, dense layers, Dropout, and a final classification layer. These additions were designed to reduce spatial dimensions, capture relevant features, learn complex patterns, and prevent overfitting.

To ensure robust evaluation, we used 5-fold cross-validation on the 80% training set, while the 20% holdout set was reserved for additional validation during training (Yadav & Shukla, 2016). This approach provided a reliable framework for both binary and multi-class classification tasks, ensuring effective training and validation for the models.

*2.3.4 Classification Using Machine Learning Based Methods*

We selected ML algorithms including Extreme Gradient Boosting (XGB) (T. Chen et al., 2024), Random Forest (RF) (Biau & Scornet, 2016), k-Nearest Neighbors (k-NN) (Larose & Larose, 2014), Support Vector Machine (SVM) (Schölkopf, 1998), and

Multi-Layer Perceptron (MLP) (Chan et al., 2023) for their proven effectiveness in classification tasks, including skin lesion classification. XGB combines the predictive power of decision trees with the efficiency of a gradient boosting framework, making it particularly effective for skin lesion classification. RF's ensemble learning approach, which integrates multiple decision trees, ensures robust and accurate predictions. k-NN, which classifies data points based on their proximity to neighbors, excels in capturing local patterns. SVM is highly versatile, adapting to both linear and nonlinear tasks, and performs well in high-dimensional feature spaces, making it suitable for the complexity of skin lesion data. MLP, with its interconnected neural network structure, learns patterns by adjusting its connections, making it an effective tool for skin lesion classification.

Before training the ML models on both binary and multi-classification tasks, data preparation included a choice between applying data augmentation or SMOTE. These preprocessing steps were essential to address class imbalance and enhance feature diversity. Feature extraction focused on the ABCD rule, extracting color, shape, and texture features, which provided comprehensive representations of skin lesions. For feature engineering, PCA was applied to reduce dimensionality while retaining the most relevant features. To ensure robust evaluation and mitigate overfitting, a 5-fold cross-validation technique (Yadav & Shukla, 2016) was employed. This method divided the dataset into five subsets, using each subset as a validation set once, while the remaining four subsets were used for training. This approach provided a comprehensive assessment of the model's performance by validating it on multiple data partitions, ensuring reliable and generalizable results.

## 3. Results

### 3.1 Performance Evaluation of the Brain MRI Tissue Segmentation Results

To assess the performance of the models for the multiclass segmentation task, several evaluation metrics were computed on the validation set, including Dice Coefficient (DSC) (Bertels et al., 2019), Hausdorff Distance (HD) (Huttenlocher et al., 1993), and Average Volumetric Difference (AVD). The DSC quantified the spatial overlap between the predicted and ground truth segmentations, with higher scores indicating better segmentation accuracy. HD measured the maximum distance between the boundaries of the predicted and ground truth segmentations, providing insights into spatial dissimilarities. AVD calculated the average absolute volume difference between the predicted and ground truth segmentations, offering an overall measure of volume agreement. These metrics collectively evaluated the accuracy and reliability of the segmentation results. Tables 1 and 2 present the performance results of the deep learning-based and Atlas-based methods respectively.

**Table 1**

**Result of the deep learning methods for the brain tissue segmentation task.** The nnU-Net achieved the best overall segmentation performance across CSF, GM, and WM segmentation tasks.

| MODEL (Backbone) | DATA EXTRACTION | CLASS | MEAN DSC | MEAN HD | MEAN AVD |
|---|---|---|---|---|---|
| 2D U-Net | Patches | CSF | 0.737812 | 28.307013 | -0.134912 |
| | | GM | 0.858232 | 14.177120 | -0.115652 |
| | | WM | 0.838790 | 12.374397 | 0.275064 |
| | | **AVERAGE** | **0.811611** | **18.286177** | **0.175209** |
| 2D U-Net (ResNet34) | Patches | CSF | 0.664457 | 41.240239 | 0.447861 |
| | | GM | 0.820576 | 16.725408 | -0.168499 |
| | | WM | 0.812676 | 12.327939 | 0.301366 |
| | | **AVERAGE** | **0.765903** | **23.431195** | **0.305909** |
| 2D U-Net (InceptionResnetv2) | Patches | CSF | 0.653226 | 41.583212 | 0.330947 |
| | | GM | 0.806399 | 15.079069 | -0.201626 |
| | | WM | 0.802612 | 11.868298 | 0.383850 |
| | | **AVERAGE** | **0.754079** | **22.843526** | **0.305474** |
| 2D U-Net | 2D Slices | CSF | 0.775183 | 18.396628 | 0.118932 |
| | | GM | 0.834245 | 8.879626 | -0.108875 |
| | | WM | 0.825270 | 11.650975 | 0.397680 |
| | | **AVERAGE** | **0.811566** | **15.522586** | **0.208496** |
| 2D U-Net (ResNet34) | 2D Slices | CSF | 0.849564 | 17.430012 | 0.049474 |
| | | GM | 0.899271 | 9.127219 | 0.024360 |
| | | WM | 0.902502 | 10.018658 | 0.108416 |

|                     |            | AVERAGE | 0.883779 | 12.191963 | 0.06075  |
|---------------------|------------|---------|----------|-----------|----------|
| 2D U-Net (ResNet50) | 2D Slices  | CSF     | 0.841105 | 31.044868 | 0.125596 |
|                     |            | GM      | 0.902972 | 12.265630 | 0.076598 |
|                     |            | WM      | 0.908532 | 11.595280 | 0.022717 |
|                     |            | AVERAGE | 0.884203 | 18.301926 | 0.074970 |
| 3D U-Net (ResNet34) | 3D Volume  | CSF     | 0.854851 | 40.551013 | 0.047525 |
|                     |            | GM      | 0.915938 | 13.640384 | 0.062216 |
|                     |            | WM      | 0.912901 | 12.651255 | 0.065367 |
|                     |            | AVERAGE | 0.894563 | 22.280884 | 0.058369 |
| Monai U-Net         | 3D Volume  | CSF     | 0.728873 | 23.148181 | 0.162052 |
|                     |            | GM      | 0.767101 | 11.140751 | 0.209706 |
|                     |            | WM      | 0.871578 | 9.780009  | 0.087166 |
|                     |            | AVERAGE | 0.789184 | 14.689647 | 0.152975 |
| nnU-Net             | 3D Volume  | CSF     | 0.923918 | 9.877988  | 0.053841 |
|                     |            | GM      | 0.949979 | 8.322139  | 0.017246 |
|                     |            | WM      | 0.945194 | 6.735176  | 0.042588 |
|                     |            | AVERAGE | 0.939697 | 8.311768  | 0.037892 |

**Table 2**

**Result of the atlas-based methods for the brain tissue segmentation task.** The multi-Atlas methods outperformed the Probabilistic Atlas methods in terms of mean DSC.

| ATLAS                | SEGMENTATION METHOD            | CLASS   | MEAN DSC  |
|----------------------|--------------------------------|---------|-----------|
| Probabilistic Atlas  | Label Propagation              | CSF     | 0.66326   |
|                      |                                | GM      | 0.77754   |
|                      |                                | WM      | 0.71004   |
|                      |                                | AVERAGE | 0.716947  |
| Probabilistic Atlas  | Tissue Model                   | CSF     | 0.11978   |
|                      |                                | GM      | 0.59208   |
|                      |                                | WM      | 0.58324   |
|                      |                                | AVERAGE | 0.4317    |
| Multi-Atlas          | Majority Voting                | CSF     | 0.67904   |
|                      |                                | GM      | 0.79288   |
|                      |                                | WM      | 0.7082    |
|                      |                                | AVERAGE | 0.7267    |
| Multi-Atlas          | Weighted by Mutual Information | CSF     | 0.67164   |
|                      |                                | GM      | 0.78972   |
|                      |                                | WM      | 0.71514   |
|                      |                                | AVERAGE | 0.7255    |

From the results, the nnU-Net achieved the best overall segmentation performance which operates on 3D volumes. This indicates that the ability to process full 3D data significantly enhances spatial feature learning, leading to higher segmentation accuracy compared to 2D approaches. Additionally, the 2D U-Net model performed similarly across patches and slices, with slightly better results on slices. This suggests that slice-based training preserves anatomical context more effectively and highlights the limitations of patch-based training in capturing global anatomical structures. The multi-Atlas methods with Majority Voting and Weighted by Mutual Information equally outperformed the Probabilistic Atlas method. This demonstrates the fact that multi-atlas approaches benefit from leveraging multiple atlases to improve label fusion accuracy. Among the Probabilistic Atlas methods, the Label Propagation approach showed better performance compared to the Tissue Model, suggesting that direct label transfer gains from precise registration between training and test images. Conversely, the Probabilistic Atlas Tissue Model low performance is likely due to its reliance on statistical models that are less adaptive to individual anatomical variations. Figure 13 shows the qualitative results of the brain tissue segmentation of the best models.

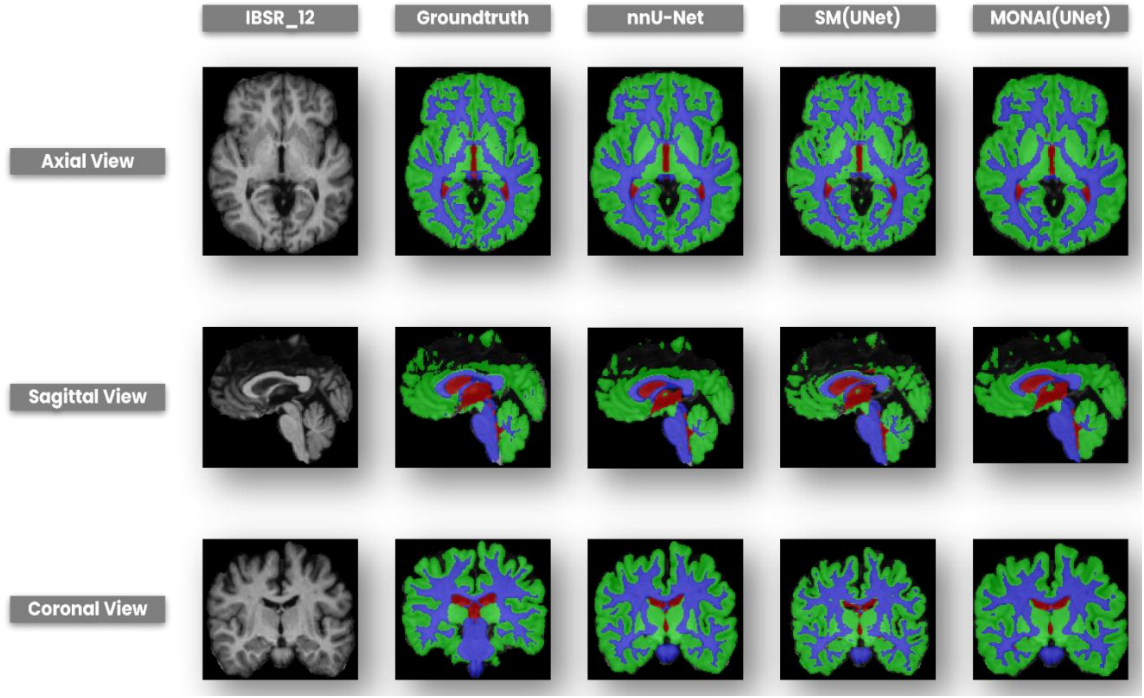

**Fig. 13.** Qualitative results of the deep learning segmentation

*3.2 Performance Evaluation of the COPD Lung CT Image Registration Results*

For the registration task, the performance of the Elastix/Transformix frameworks and pre-trained deep learning models was evaluated using Target Registration Error (TRE). TRE was computed on both the training and validation sets, quantifying the discrepancy between corresponding points in the fixed (inhale) and transformed moving (exhale) images. It served as a direct measure of registration accuracy, with smaller TRE values indicating superior alignment. The computation involved identifying anatomical landmarks or fiducial markers in both the fixed and moving images and calculating the Euclidean distance between corresponding points post-registration. Tables 3 and 4 present the performance results of the deep learning-based and Elastix/Transformix methods respectively.

While HighResNet performed slightly better than SegResNet in table 3, their results are closely comparable. HighResNet achieved a lower average TRE (7.4037 mm) compared to SegResNet (8.0161 mm), indicating slightly superior performance but with marginal differences between the two models. The best parameter configuration, which achieved the lowest average TRE (6.68 ± 5.98 mm), was a combination of multiple optimized parameters (Par0011 Affine + customized Bspline + Bspline) seen in table 4. This highlights the advantage of optimizing multiple parameters, including Affine and B-spline transformations, to achieve robust and accurate registration results. Similarly, the Par0011 Affine + customized Bspline configuration performed exceptionally well, with an average TRE of 6.85 ± 5.92 mm, close to the best-performing configuration. This emphasizes the importance of task-specific parameter customizations in enhancing registration accuracy. On the other hand, the Par0049 and Par0007 parameter sets exhibited higher TRE values, indicating limitations in their adaptability to our datasets. The Inbuilt Affine + Bspline configuration showed an average TRE of 12.91 ± 6.19 mm, suggesting that generic parameter configurations may not perform optimally for complex cases. Figures 14 and 15 provide qualitative illustrations of the best registration methods, further showcasing their effectiveness.

Table 3

**Registration results for the two deep learning models.** The values represent the mean TRE in millimetres (mm) of the registration performance. The HighResNet performed better than the SegResNet.

| MODEL | COPD1 | COPD2 | COPD3 | COPD4 | Average |
|---|---|---|---|---|---|
| SegResNet | **7.2533** | **10.2973** | 6.0914 | 8.4222 | 8.0161 |
| HighResNet | 8.2021 | 11.3182 | **3.9731** | **6.1212** | **7.4037** |

Table 4

**Registration results for various parameter configurations using Elastix and Transformix.** The values represent the mean TRE and standard deviation in millimetres (mm) of the registration performance for different COPD datasets. The "Combining all best parameters" configuration achieved the lowest average TRE (6.68 ± 5.98 mm), demonstrating the advantage of optimizing multiple parameters.

| PARAMETER(S) | COPD1 | COPD2 | COPD3 | COPD4 | Average |
|---|---|---|---|---|---|
| Inbuilt Affine + Bspline | 12.69 ± 5.94 | 18.50 ± 7.48 | 5.14 ± 4.11 | 15.31 ± 7.24 | 12.91 ± 6.19 |
| Par0049 | 18.72 ± 9.68 | 18.10 ± 8.13 | 4.52 ± 4.34 | 17.45 ± 6.98 | 14.70 ± 7.28 |
| Par0011 | 9.43 ± **5.43** | 15.86 ± 8.06 | 4.29 ± 3.48 | 10.59 ± 5.00 | 10.04 ± 5.50 |
| Inbuilt Affine + customized Bspline | 7.51 ± 6.43 | 8.19 ± 6.45 | 3.15 ± 3.62 | 15.29 ± 11.13 | 8.54 ± 6.91 |
| Par0011 Affine + customized Bspline | 7.31 ± 6.39 | 8.43 ± 6.93 | 3.08 ± 3.50 | **8.58** ± 6.85 | 6.85 ± 5.92 |
| Customized Bspline | 8.03 ± 6.60 | 8.61 ± **6.40** | 3.14 ± 3.57 | 18.00 ± 12.23 | 9.45 ± 7.2 |
| Par0007 | 10.97 ± 5.85 | 14.57 ± 6.64 | 5.13 ± **3.38** | 12.02 ± **4.30** | 10.67 ± **5.04** |
| Customized Bspline, and Bspline | 7.63 ± 6.66 | 8.51 ± 6.63 | 2.99 ± 3.45 | 17.86 ± 12.11 | 9.25 ± 7.21 |
| Combining all best parameters | **6.87** ± 6.39 | **8.33** ± 7.29 | **2.94** ± 3.41 | 8.59 ± 6.82 | **6.68** ± **5.98** |

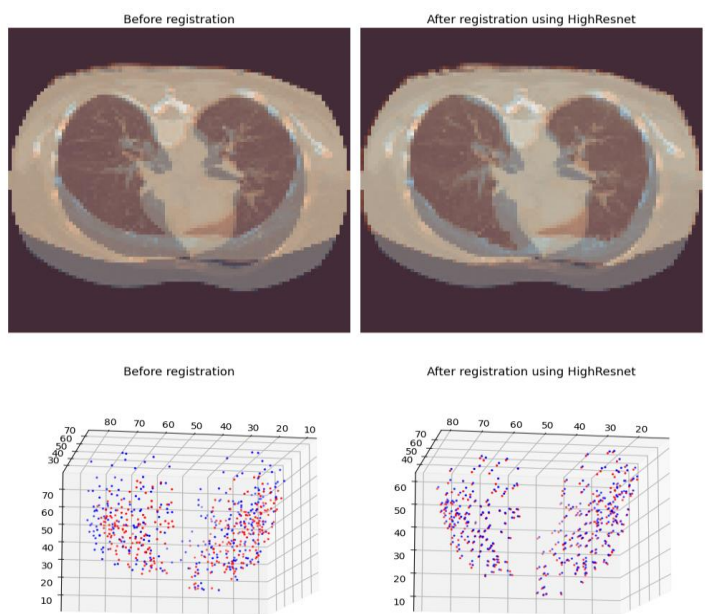

**Fig. 14.** Visualization of the result of the HighResNet model for both images and landmarks using overlay for before and after registration.

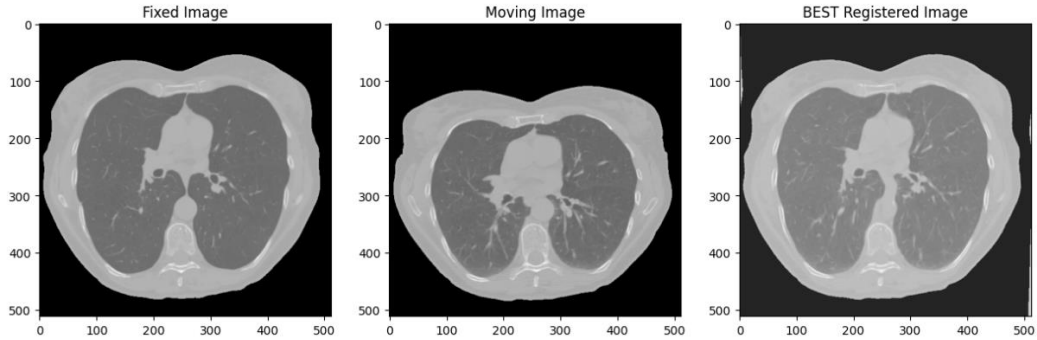

**Fig. 15.** Visualization of the result of the best parameter for the Elastix/Transformix approach

*3.3 Performance Evaluation of the Skin Lesion Classification Results*

Finally, for the classification tasks, a range of performance metrics was used to evaluate model effectiveness. These included accuracy (ACC), precision (PRC), recall (REC), F1-score, area under the curve (AUC), balanced accuracy (BACC), receiver operating characteristic (ROC) curve, and Cohen's Kappa (Kappa) (Tharwat, 2018)(Vujović, 2021). PRC measured the accuracy of positive predictions, REC assessed the model's ability to identify positive instances, and the F1-score provided a balanced evaluation of precision and recall. Balanced multiclass accuracy (BMA) was also used to account for dataset imbalance, representing the average accuracy across all classes. Cohen's Kappa measured inter-rater agreement, adjusting for chance agreement and offering a robust metric for classification performance. These metrics provided a comprehensive assessment of the model's capabilities in binary and multiclass classification tasks.

The results are divided into two groups: those for the two-class problem and those for the three-class problem. Tables 5 and 6 present the performance results for the two-class problem using deep learning-based and machine learning methods, respectively. Similarly, Tables 7 and 8 display the performance results for the three-class problem for deep learning-based and machine learning methods, respectively. These tables provide a clear comparison of model performance across binary and multiclass classification tasks, offering valuable insights into the strengths and limitations of each method.

The ensemble of RNET and IRNET outperformed all other approaches as shown in table 5b, achieving the highest metrics across the board, including ACC (0.9044), PRC (0.9145), and F1-score (0.9013). This underscores the synergy of combining these two powerful architectures. Individually, IRNET and RNET demonstrated excellent performance, highlighting their robustness and the value of ensemble methods in maximizing classification accuracy. Conversely, MNET exhibited the lowest performance among DL models, with an ACC of 0.8591, reflecting its trade-off between computational efficiency and classification accuracy.

For the ML, the ensemble of the XGB and RF models achieved results comparable to MLP shown in table 6b, demonstrating the advantages of combining ML approaches. The strong performance of MLP is unsurprising, given its close ties to deep learning methodologies. On the other hand, k-NN consistently lagged other models across all metrics, indicating that it may not be well-suited for skin cancer classification tasks. Figure 16 displays the ROC curves for the best-performing classifiers in the two-class problem, for ML and DL approach.

**Table 5**

**5a. Cross-validation (5-fold) results from two-class problem for DL approach.** RNET had the highest average cross-validation score (0.8808), followed closely by IRNET (0.8787), reflecting their strength in feature extraction and classification amongst other models used.

| FOLDS | DNET | RNET | MNET | IRNET |
|---|---|---|---|---|
| Onefold | 0.8326 | 0.8877 | 0.8298 | 0.8956 |
| Twofold | 0.8375 | 0.8873 | 0.7993 | 0.8968 |
| Threefold | 0.8214 | 0.8712 | 0.8573 | 0.8774 |
| Fourfold | 0.8383 | 0.8856 | 0.8215 | 0.8926 |
| Fivefold | 0.8429 | 0.8721 | 0.8425 | 0.8309 |
| **Average** | **0.8346** | **0.8808** | **0.8300** | **0.8787** |

**5b. Performance evaluation of the two-class problem for DL approach.** The ensemble of RNET and IRNET models outperformed all other approaches, achieving the highest metrics across the board, including ACC (0.9044), PRC (0.9145), and F1-score (0.9013). This shows the strength of ensemble methods.

| METRICS | RNET + IRNET | DNET | RNET | MNET | IRNET |
|---|---|---|---|---|---|
| ACC | 0.9044 | 0.8419 | 0.8936 | 0.8591 | 0.9015 |
| PRC | 0.9145 | 0.8410 | 0.8938 | 0.8778 | 0.9162 |
| REC | 0.8885 | 0.8365 | 0.8890 | 0.8284 | 0.8799 |
| F1-score | 0.9013 | 0.8387 | 0.8914 | 0.8524 | 0.8977 |
| AUC | 0.9041 | 0.8418 | 0.8935 | 0.8585 | 0.9011 |

**Table 6**

**6a. Cross-validation (5-fold) results from two-class problem for ML approach.** XGB and MLP consistently achieved the highest average cross-validation scores among individual ML models with an average score of 0.81 across the fivefold.

| FOLDS | XGB | RF | k-NN | SVM | MLP |
|---|---|---|---|---|---|
| Onefold | 0.8009 | 0.7937 | 0.7654 | 0.7924 | 0.8009 |
| Twofold | 0.8161 | 0.8180 | 0.7861 | 0.8072 | 0.8213 |
| Threefold | 0.8170 | 0.8154 | 0.7825 | 0.7990 | 0.8138 |
| Fourfold | 0.8240 | 0.8230 | 0.7855 | 0.8164 | 0.8184 |
| Fivefold | 0.8128 | 0.8111 | 0.7904 | 0.8141 | 0.8144 |
| **Average** | **0.8142** | **0.8122** | **0.7820** | **0.8058** | **0.8138** |

**6b. Performance evaluation of the two-class problem for ML approach.** The ensemble of XGB and RF shows comparative performance with the MLP.

| METRICS | XGB + RF | XGB | RF | k-NN | SVM | MLP |
|---|---|---|---|---|---|---|
| ACC | 0.8301 | 0.8232 | 0.8180 | 0.7895 | 0.8138 | 0.8327 |
| PRC | 0.8273 | 0.8227 | 0.8099 | 0.7759 | 0.8120 | 0.8296 |
| REC | 0.8268 | 0.8161 | 0.8225 | 0.8038 | 0.8080 | 0.8300 |
| F1-score | 0.8270 | 0.8194 | 0.8162 | 0.7896 | 0.8100 | 0.8298 |
| AUC | 0.8300 | 0.8231 | 0.8180 | 0.7898 | 0.8137 | 0.8327 |
| BACC | 0.8300 | 0.8231 | 0.8180 | 0.7898 | 0.8137 | 0.8327 |

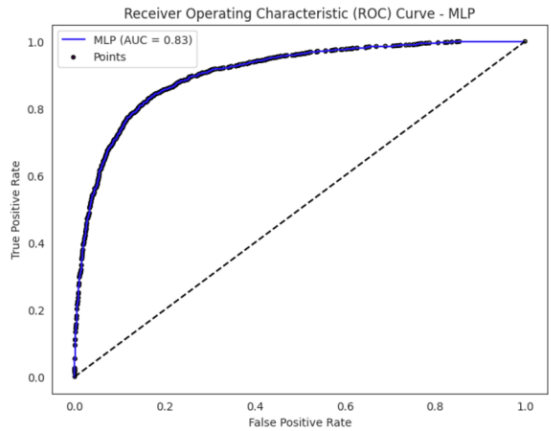
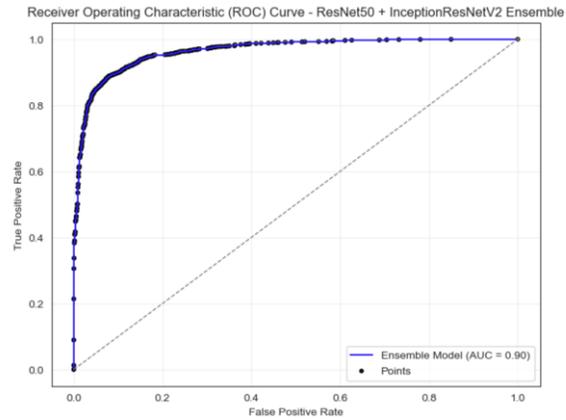

16(a)            16(b)

**Fig. 16.** ROC Curves for best performing classifiers in binary classification: (a)MLP (b) ResNet50 + InceptionResNetV2.

In the three-class problem, the combined RNET and IRNET model also outperformed all individual models as represented in table 7b, highlighting the benefits of ensembling for enhanced accuracy and robustness in multi-class classification. Additionally, based on the strong performance of IRNET, it was chosen for training the one-vs-all classification approach in table 7d. Particularly for melanoma and BCC, IRNET demonstrated significant adaptability to different classification setups, achieving high performance across metrics. However, SCC classification proved to be the most challenging due to its imbalanced representation in the dataset, as evidenced by the lower REC and F1-score for SCC vs. Others. Figure 17 illustrates the ROC curves for the one-vs-all results using IRNET

**Table 7**

**7a. Cross-validation (5-fold) results from three-class problem for DL approach.** RNET achieved the highest average cross-validation score, closely followed by IRNET.

| FOLDS | DNET | RNET | MNET | IRNET |
|---|---|---|---|---|
| Onefold | 0.8397 | 0.9536 | 0.8226 | 0.9355 |
| Twofold | 0.8558 | 0.9405 | 0.8609 | 0.9607 |
| Threefold | 0.8700 | 0.9506 | 0.8246 | 0.9425 |
| Fourfold | 0.8266 | 0.9506 | 0.8145 | 0.9516 |
| Fivefold | 0.7510 | 0.9415 | 0.8266 | 0.9294 |
| Average | **0.8286** | **0.9474** | **0.8298** | **0.9439** |

**7b. Performance evaluation of the three-class problem for DL approach.** The ensemble model RNET and IRNET achieved the best performance across all metrics.

| METRICS | RNET + IRNET | DNET | RNET | MNET | IRNET |
|---|---|---|---|---|---|
| ACC | 0.9362 | 0.8094 | 0.9157 | 0.8378 | 0.9260 |
| PRC | 0.9368 | 0.8311 | 0.9181 | 0.8521 | 0.9270 |
| REC | 0.9362 | 0.8094 | 0.9157 | 0.8378 | 0.9260 |
| F1-score | 0.9357 | 0.8169 | 0.9155 | 0.8437 | 0.9257 |
| BMA | 0.8882 | 0.7647 | 0.8655 | 0.7643 | 0.8802 |
| KAPPA | **0.8841** | **0.6672** | **0.8472** | **0.7147** | **0.8658** |

**7c. Cross-validation (5-fold) results from three-class problem for DL approach (One-vs-All Results using IRNET).**

| FOLDS | Melanoma vs. Others | BCC vs. Others | SCC vs. Others |
|---|---|---|---|
| Onefold | 0.9420 | 0.9125 | 0.9371 |
| Twofold | 0.9479 | 0.9381 | 0.9695 |
| Threefold | 0.9547 | 0.9183 | 0.9518 |
| Fourfold | 0.9390 | 0.9350 | 0.9656 |
| Fivefold | 0.9547 | 0.9320 | 0.9646 |
| Average | **0. 9477** | **0. 9272** | **0. 9577** |

**7d. Performance evaluation of the three-class problem for DL approach (One-vs-All Results using IRNET).** Despite the performance of SCC vs. Others in accuracy, it has low REC as a result of class imbalance.

| METRICS | Melanoma vs. Others | BCC vs. Others | SCC vs. Others |
|---|---|---|---|
| ACC | **0.9465** | **0.9535** | **0.9693** |
| PRC | 0.9320 | 0.9660 | 0.9104 |
| REC | 0.9705 | 0.9137 | 0.6489 |
| F1-score | 0.9509 | 0.9391 | 0.7578 |
| AUC | 0.9447 | 0.9465 | 0.8219 |
| BACC | 0.9447 | 0.9465 | 0.8219 |

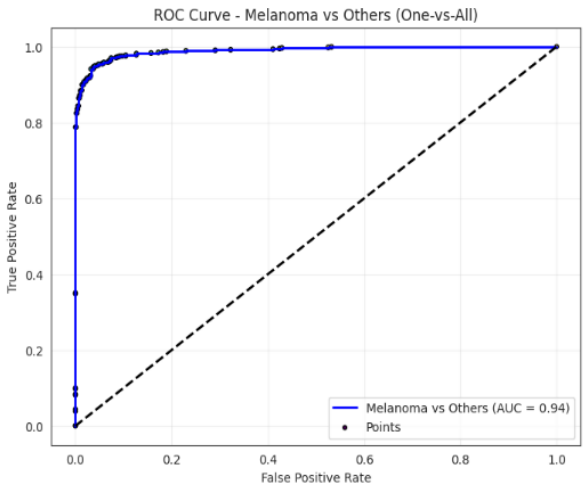

17(a)

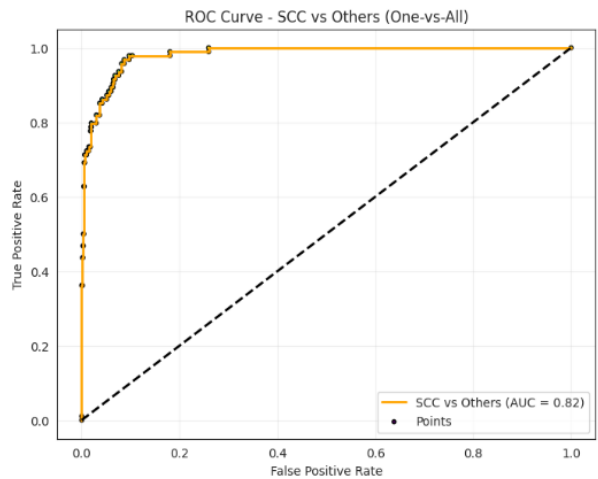

17(b)

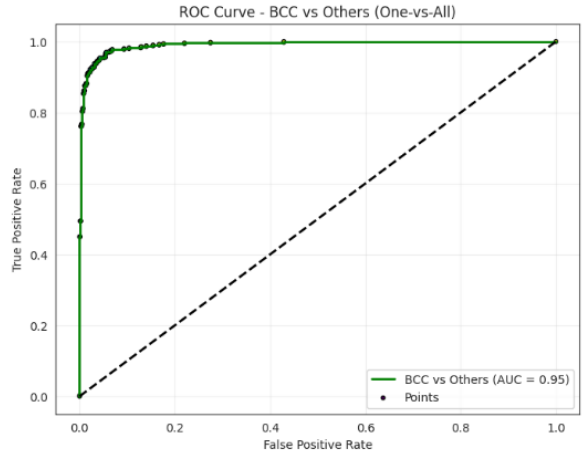

17(c)

**Fig. 17.** ROC Curve for the One-vs-All Results using IRNET (a) Melanoma vs. Others, (b) BCC vs. Others, and (c) SCC vs. Others.

For the ML in three class problem, the combination of XGB and RF consistently demonstrated strong performance; however, MLP outperformed all models overall in both setups (SMOTE and Augmentation) seen in table 8b and 8c. Particularly, MLP achieved slightly better KAPPA results using SMOTE, which could be attributed to the fact that SMOTE directly address class imbalance by synthetically expanding minority class samples, ensuring a more balanced representation during training. In contrast, augmentation enhances data diversity but may not fully mitigate class imbalance, particularly in datasets with severely underrepresented classes like in our case. This potentially explain why SMOTE resulted in slightly better performance. Figure 18 illustrates the ROC curves of the ensemble model (RF and XGB) and MLP across both setups.

**Table 8**

**8a. Cross-validation (5-fold) results from three-class problem for ML approach (with SMOTE).**

| FOLDS | XGB + RF | XGB | RF | k-NN | SVM | MLP |
|---|---|---|---|---|---|---|
| Onefold | 0.8552 | 0.8423 | 0.8399 | 0.7369 | 0.7289 | 0.8632 |
| Twofold | 0.8640 | 0.8600 | 0.8504 | 0.7377 | 0.7450 | 0.8552 |
| Threefold | 0.8890 | 0.8930 | 0.8665 | 0.7481 | 0.7418 | 0.8785 |
| Fourfold | 0.8744 | 0.8696 | 0.8688 | 0.7512 | 0.7576 | 0.8688 |
| Fivefold | 0.8800 | 0.8776 | 0.8816 | 0.7633 | 0.7448 | 0.8696 |
| Average | **0.8725** | **0.8685** | **0.8614** | **0.7474** | **0.7436** | **0.8671** |

**8b. Performance evaluation of the three-class problem for ML approach (with SMOTE).** The MLP classifier achieved the overall best results using SMOTE.

| METRICS | XGB + RF | XGB | RF | k-NN | SVM | MLP |
|---|---|---|---|---|---|---|
| ACC | 0.8417 | 0.8323 | 0.8268 | 0.7252 | 0.7819 | 0.8504 |
| PRC | 0.8341 | 0.8256 | 0.8192 | 0.7760 | 0.8005 | 0.8514 |
| REC | 0.8417 | 0.8323 | 0.8268 | 0.7252 | 0.7819 | 0.8504 |
| F1-score | 0.8364 | 0.8282 | 0.8186 | 0.7417 | 0.7894 | 0.8506 |
| BMA | 0.7066 | 0.7021 | 0.6819 | 0.6867 | 0.7142 | 0.7558 |
| KAPPA | **0.6781** | **0.6781** | **0.6781** | **0.5343** | **0.6187** | **0.7315** |

**8c. Cross-validation (5-fold) results from three-class problem for ML approach (with Augmentation).**

| FOLDS | XGB + RF | XGB | RF | k-NN | SVM | MLP |
|---|---|---|---|---|---|---|
| Onefold | 0.8234 | 0.8242 | 0.8202 | 0.7470 | 0.7757 | 0.8417 |
| Twofold | 0.8449 | 0.8465 | 0.8353 | 0.7534 | 0.7979 | 0.8520 |
| Threefold | 0.8551 | 0.8591 | 0.8519 | 0.7635 | 0.7978 | 0.8527 |
| Fourfold | 0.8376 | 0.8432 | 0.8304 | 0.7333 | 0.7763 | 0.8272 |
| Fivefold | 0.8376 | 0.8384 | 0.8352 | 0.7675 | 0.7906 | 0.8694 |
| Average | **0.8397** | **0.8423** | **0.8346** | **0.7529** | **0.7877** | **0.8486** |

**8d. Performance evaluation of the three-class problem for ML approach (with Augmentation).** Again, the MLP classifier achieved the overall best results using Augmentation

| METRICS | XGB + RF | XGB | RF | k-NN | SVM | MLP |
|---|---|---|---|---|---|---|
| ACC | 0.8386 | 0.8354 | 0.8165 | 0.7646 | 0.7866 | 0.8465 |
| PRC | 0.8460 | 0.8295 | 0.8227 | 0.7464 | 0.7746 | 0.8483 |
| REC | 0.8386 | 0.8354 | 0.8165 | 0.7646 | 0.7866 | 0.8465 |
| F1-score | 0.8216 | 0.8214 | 0.7952 | 0.7529 | 0.7625 | 0.8472 |
| BMA | 0.6491 | 0.6559 | 0.6134 | 0.5875 | 0.5760 | 0.7593 |
| KAPPA | **0.6537** | **0.6537** | **0.6537** | **0.5641** | **0.6017** | **0.7250** |

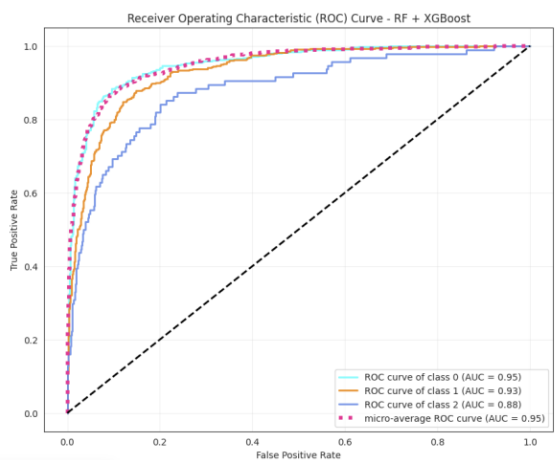

18(a)

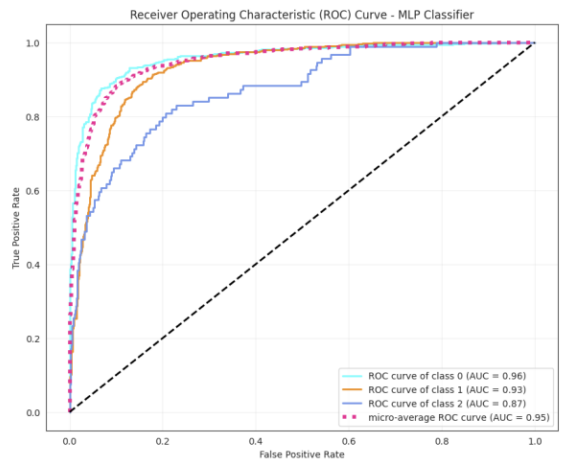

18(b)

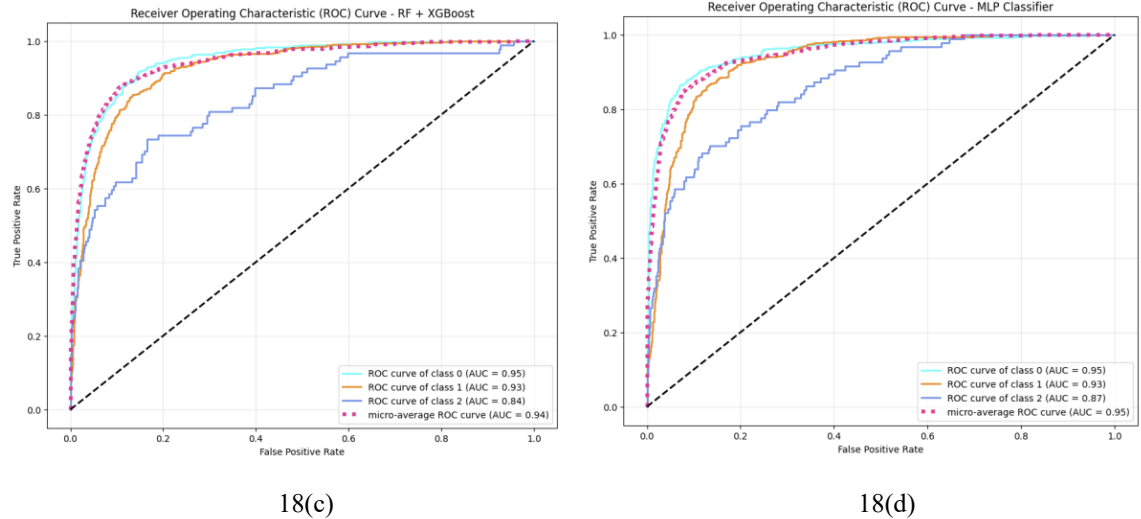

18(c)                               18(d)

**Fig. 17.** ROC Curves for best performing classifiers in multi-class classification: (a) RF + XGBoost (with SMOTE), (b) MLP (with SMOTE), (c) RF + XGBoost (with Augmentation), and (d) MLP (with Augmentation).

## 4. Discussion

*4.1 Discussion of the Brain MRI Tissue Segmentation Results*

The segmentation results highlight the superior performance of 3D deep learning models, with nnU-Net achieving the best overall results across CSF, GM, and WM segmentation tasks. Its ability to process full 3D volumes allowed it to effectively capture spatial relationships and anatomical structures, achieving a remarkable average Dice Coefficient (DSC) of 0.9397, significantly outperforming other methods. 3D U-Net on ResNet34 backbone, also operating on 3D volumes, followed closely, further demonstrating the importance of leveraging 3D data for spatial feature learning. In contrast, the 2D slice-based approaches outperformed patch-based methods, with the ResNet50 slice-based model achieving an average DSC of 0.8842. This suggests that maintaining anatomical context during training significantly influences segmentation outcomes, and while patch-based methods are less computationally demanding, they face inherent limitations in preserving global structural information.

Atlas-based methods offered a robust alternative, with multi-Atlas approaches surpassing Probabilistic Atlas techniques in segmentation accuracy using DSC. Methods like Majority Voting and Weighted by Mutual Information leveraged the collective strength of multiple atlases, achieving average DSCs of 0.7267 and 0.7255, respectively. Within the Probabilistic Atlas group, the Label Propagation method outperformed the Tissue Model, reflecting the importance of accurate registration for effective label transfer. The underperformance of the Tissue Model signifies the limitations of relying on generalized statistical assumptions, particularly when dealing with individual anatomical variability. These results underscore the importance of selecting segmentation methods based on the data's complexity and task requirements, with 3D deep learning models leading the way, and multi-atlas approaches serving as robust alternatives for scenarios where deep learning is not feasible.

*4.2 Discussion of the COPD Lung CT Image Registration Results*

The registration results provide a clear comparison between classical Elastix-based approaches and deep learning (DL) models, highlighting their distinct strengths and limitations. The Elastix framework, with its flexible parameter configurations, demonstrated strong performance, particularly with optimized settings. The combination of the best parameters (Par0011 Affine, customized B-spline, and B-spline) achieved the lowest average TRE of 6.68 ± 5.98 mm, emphasizing the critical role of tailoring parameter maps to specific datasets. Similarly, the Par0011 Affine + customized B-spline configuration delivered comparable results, with an average TRE of 6.85 ± 5.92 mm. These findings underscore the potential of systematic parameter optimization in classical methods to enhance registration accuracy, especially for aligning inhale and exhale lung CT images. Conversely, generic configurations like Inbuilt Affine and Bspline struggled with complex cases, as reflected in a higher average TRE of 12.91 ± 6.19 mm, highlighting the need for task-specific parameter customization.

Deep learning models brought a dynamic and adaptive dimension to the registration task. HighResNet and SegResNet showed closely comparable performance, with HighResNet slightly outperforming SegResNet, achieving an average TRE of 7.4037 mm compared to 8.0161 mm. HighResNet's strength lies in its ability to handle high-resolution data effectively, while SegResNet offered a balanced trade-off between precision and computational efficiency. Despite their solid performance, the DL models did not surpass the best-optimized Elastix configurations, demonstrating the continued relevance of classical methods when paired with robust parameter tuning. However, DL models provide the advantage of scalability and adaptability, which can be further enhanced with larger datasets.

Overall, the results highlight the complementary strengths of classical and DL-based registration methods. Elastix excels in scenarios requiring flexibility and precision, especially when computational resources are limited, and parameter optimization is feasible. On the other hand, DL models showcase strong adaptability and scalability, making them potentially promising for future advancements, and cases with larger datasets. Choosing the appropriate registration methodology should depend on the specific requirements of the task, balancing accuracy, resource availability, and the ability to handle dataset variability.

*4.3 Discussion of the Skin Lesion Classification Results*

The results from these tasks reveal distinct strengths and trade-offs between ML and DL approaches across binary, multi-class, and One-vs-All setups. In the binary classification task (Nevus vs. Others), DL models like InceptionResNetV2B0 (IRNET) and ResNet50 (RNET) demonstrated superior performance, with accuracies of 90.15% and 89.36%, respectively, and their ensemble achieving 90.44%. In comparison, the best-performing ML model, Multi-Layer Perceptron (MLP), achieved 83.27%, closely followed by the ensemble of Random Forest (RF) and XGBoost (XGB) at 83.01%. These findings highlight the capability of DL models to capture complex features and achieve higher accuracies, particularly when combined into ensembles. However, ML approaches remain competitive, offering respectable performance with lower computational demands, faster training times and interpretability as the features are handcrafted.

In the three-class classification task (melanoma, BCC, SCC), the incorporation of SMOTE effectively improved the balance across class performances for ML models, with MLP achieving the highest accuracy of 85.04% and a KAPPA of 73.15%. This shows the importance of addressing class imbalance, especially for underrepresented categories like SCC. Among DL models, IRNET consistently outperformed other single models, achieving an accuracy of 92.60% and a KAPPA of 86.58%. However, the ensemble of RNET and IRNET achieved the overall best performance, with an accuracy of 93.62% and a KAPPA of 88.41%. Using IRNET in the One-vs-All configurations yielded accuracies of 94.65% for melanoma, 95.35% for BCC, and 96.93% for SCC, further demonstrating its robustness. These results show that while DL models dominate in accuracy, ML models, when enhanced with effective preprocessing techniques like SMOTE, provide robust and interpretable alternatives.

Overall, the comparison between DL and ML approaches underscores their complementary advantages. DL models excel in accuracy, leveraging transfer learning and deep hierarchical features, making them ideal for large-scale applications requiring high precision. However, they demand significant computational resources and are less interpretable in this context. On the other hand, ML models, augmented by feature engineering and imbalance correction techniques, offer interpretable and resource-efficient solutions, particularly suitable for clinical settings with constrained resources. These findings emphasize the importance of aligning classification methodologies with the specific requirements of the task, balancing accuracy, interpretability, and resource availability.

## 5. Conclusion

This study demonstrates the potential of combining classical and deep learning approaches to address three key tasks in medical imaging, including segmentation, registration, and classification. Each task highlighted the distinct advantages and trade-offs of various methods, underscoring the importance of selecting techniques based on the specific requirements and constraints of the problem. In brain tissue segmentation task, 3D deep learning models like nnU-Net outperformed other methods by leveraging their ability to process full volumetric data. Slice-based 2D models showed competitive results, highlighting their effectiveness in scenarios where computational resources are constrained. Atlas-based segmentation methods, particularly multi-Atlas approaches, provided robust alternatives, offering reliable results when deep learning is not feasible, although they are limited by their dependence on accurate registration and effective label fusion techniques.

For lung CT registration, classical Elastix-based methods with optimized parameters achieved the best overall accuracy, demonstrating the power of parameter tuning in aligning inhale and exhale images and landmarks with high precision. While deep learning models such as HighResNet and SegResNet showed promising results, they did not surpass the best classical configurations, indicating that classical methods remain highly effective when robust parameter tuning is feasible. Finally, in the skin lesion classification task, the deep learning models such as InceptionResNetV2 and ResNet50 excelled, particularly when combined into ensembles, achieving the highest accuracy and robustness in both binary and multi-class tasks. Techniques like SMOTE effectively addressed class imbalance, significantly enhancing the performance of ML models. Also, machine learning models like MLP, while not as accurate as deep learning ensembles, provided interpretable and computationally efficient solutions, making them suitable for resource-constrained settings.

Conclusively, these results highlight the complementary strengths of classical and deep learning approaches, emphasizing the need for task-specific adaptations. Deep learning models excel in tasks requiring high accuracy and complex feature learning, while classical methods and machine learning approaches offer reliable and interpretable solutions in scenarios with limited resources or specific constraints. This study underscores the importance of balancing accuracy, computational efficiency, and interpretability to achieve optimal outcomes in medical imaging tasks for diverse applications.

**CRediT authorship contribution statement**

**DT Anyimadu:** Writing – review & editing, Writing – original draft, Data curation, Software, Visualization, Validation, Methodology, Investigation, Formal analysis. **TA Suleiman:** Writing – review & editing, Writing – original draft, Data curation, Software, Visualization, Validation, Methodology, Formal analysis. **MI Hossain:** Writing – review & editing, Data curation, Project administration, Validation, Methodology, Investigation.

**Declaration of competing interest**

The authors declare that they have no known competing financial interests or personal relationships that could have appeared to influence the work reported in this paper.